\documentclass{article}
\newcommand{\dataProcessingFinalcorpussize}{295}

\newcommand{\authorMetricsUniqueauthors}{1210}

\newcommand{\editormissingMean}{14}
\newcommand{\editormissingLow}{5}
\newcommand{\editormissingHigh}{25}
\newcommand{\editormissingGooglemean}{32}
\newcommand{\editormissingGooglelow}{18}
\newcommand{\editormissingGooglehigh}{47}
\newcommand{\editormissingMetamean}{61}
\newcommand{\editormissingMetalow}{46}
\newcommand{\editormissingMetahigh}{76}
\newcommand{\editormissingMicrosoftmean}{8}
\newcommand{\editormissingMicrosoftlow}{1}
\newcommand{\editormissingMicrosofthigh}{16}
\newcommand{\editortiesMean}{34}
\newcommand{\editortiesLow}{23}
\newcommand{\editortiesHigh}{43}
\newcommand{\editedpapertiesMean}{35}
\newcommand{\editedpapertiesLow}{27}
\newcommand{\editedpapertiesHigh}{42}
\newcommand{\authordisclosureprobBetamean}{0.161}
\newcommand{\authordisclosureprobBetalow}{-0.026}
\newcommand{\authordisclosureprobBetahigh}{0.366}
\newcommand{\authordisclosureyearsDiffmean}{6}
\newcommand{\authordisclosureyearsDifflow}{-2}
\newcommand{\authordisclosureyearsDiffhigh}{14}
\newcommand{\authordisclosureyearsTenmean}{58}
\newcommand{\authordisclosureyearsTenlow}{53}
\newcommand{\authordisclosureyearsTenhigh}{63}
\newcommand{\authorprolificprobBetalow}{0.56}
\newcommand{\authorprolificprobBetahigh}{0.87}
\newcommand{\authorprolificprobBetamean}{0.71}
\newcommand{\editorialindependenceMean}{34}
\newcommand{\editorialindependenceLow}{42}
\newcommand{\editorialindependenceHigh}{27}
\newcommand{\editorauthtiedMean}{66}
\newcommand{\editorauthtiedLow}{58}
\newcommand{\editorauthtiedHigh}{73}
\newcommand{\editortieprobabilityBetamean}{0.311}
\newcommand{\editortieprobabilityBetalow}{0.084}
\newcommand{\editortieprobabilityBetahigh}{0.589}

\newcommand{\literatureIndustryMean}{8}
\newcommand{\literatureIndustryLow}{5}
\newcommand{\literatureIndustryHigh}{12}

\newcommand{\literatureDisclosedMean}{13}
\newcommand{\literatureDisclosedLow}{10}
\newcommand{\literatureDisclosedHigh}{17}

\newcommand{\literatureIdentifiableMean}{20}
\newcommand{\literatureIdentifiableLow}{15}
\newcommand{\literatureIdentifiableHigh}{25}

\newcommand{\literatureFoundMean}{49}
\newcommand{\literatureFoundLow}{43}
\newcommand{\literatureFoundHigh}{55}

\newcommand{\affirmnolitFoundMean}{42}
\newcommand{\affirmnolitFoundLow}{36}
\newcommand{\affirmnolitFoundHigh}{49}
\newcommand{\affirmnolitAnyN}{227}

\newcommand{\authorsNonemployedgini}{0.919}
\newcommand{\authorsToptenpercentcontrol}{79.0}

\newcommand{\authorsaAnyMean}{21}
\newcommand{\authorsaAnyLow}{19}
\newcommand{\authorsaAnyHigh}{23}

\newcommand{\authorsbTwitterMean}{1}
\newcommand{\authorsbTwitterLow}{0}
\newcommand{\authorsbTwitterHigh}{2}

\newcommand{\authorsbMicrosoftMean}{6}
\newcommand{\authorsbMicrosoftLow}{5}
\newcommand{\authorsbMicrosoftHigh}{7}

\newcommand{\authorsbMultipleMean}{7}
\newcommand{\authorsbMultipleLow}{6}
\newcommand{\authorsbMultipleHigh}{8}

\newcommand{\authorsbGoogleMean}{8}
\newcommand{\authorsbGoogleLow}{7}
\newcommand{\authorsbGoogleHigh}{10}

\newcommand{\authorsbMetaMean}{14}
\newcommand{\authorsbMetaLow}{12}
\newcommand{\authorsbMetaHigh}{16}
\newcommand{\disclosuretrendBetamean}{0.01}
\newcommand{\disclosuretrendBetalow}{-0.107}
\newcommand{\disclosuretrendBetahigh}{0.133}

\newcommand{\editorbasicsEditedpapers}{167}
\newcommand{\editorcoauthorBetamean}{1.641}
\newcommand{\editorcoauthorBetalow}{0.424}
\newcommand{\editorcoauthorBetahigh}{2.95}
\newcommand{\editorcoauthorPmean}{11}
\newcommand{\editorcoauthorPlow}{5}
\newcommand{\editorcoauthorPhigh}{18}
\newcommand{\editorbasicsagainNedtiors}{80}

\newcommand{\saturationTotalpapers}{295}

\newcommand{\saturationEsitimatedmean}{79.9}
\newcommand{\saturationEstimatedlow}{74.05}
\newcommand{\saturationEstimatedhigh}{85.43}

\newcommand{\reviewerMean}{29.9}
\newcommand{\reviewerLow}{19.9}
\newcommand{\reviewerHigh}{39.9}
\newcommand{\reviewerTotal}{82}

\newcommand{\reviewerUniquepapers}{49}
\newcommand{\authorsupportyearsMean}{1.27}
\newcommand{\authorsupportyearsMeanlow}{1.05}
\newcommand{\authorsupportyearsMeanhigh}{1.55}
\newcommand{\authorsupportyearsMax}{33}
\newcommand{\totalsAltmetric}{154623}
\newcommand{\totalsDimensions}{53120}
\newcommand{\totalsPolicy}{745}
\newcommand{\totalsNews}{15708}
\newcommand{\totalsSocialmedia}{100863}
\newcommand{\totalsWikipedia}{418}
\newcommand{\totalsAvgdimensions}{180}

\newcommand{\biasbasicsLowcommunity}{31}
\newcommand{\biasbasicsHighcommunity}{82}

\usepackage{arxiv}

\usepackage[utf8]{inputenc}
\usepackage[T1]{fontenc}
\usepackage{url, booktabs, amsfonts, nicefrac, microtype}
\usepackage{graphicx, graphics, amssymb, amsmath, epsfig, color}

\usepackage{xr-hyper}                       
\usepackage[
    style=numeric-comp, 
    sorting=none, 
    backend=biber
]{biblatex}                                 %
\usepackage[hidelinks]{hyperref}             

\title{Industry Influence in High-Profile Social Media Research}

\author{
  Joseph Bak-Coleman$^{1,2,3,4,5\ast}$, Jevin West$^{6,7}$,\\
  \textbf{Cailin O'Connor}$^{8,9}$, and \textbf{Carl T. Bergstrom}$^{1,2,7}$ \\
  \vspace{2mm} \\
  \small $^1$Department of Biology, University of Washington, Seattle, WA, USA \\
  \small $^2$Santa Fe Institute, Santa Fe, NM, USA \\
  \small $^3$Centre for the Advanced Study of Collective Behavior, University of Konstanz, Germany \\
  \small $^4$Department of Collective Behavior, Max Planck Institute of Animal Behavior, Germany \\
  \small $^5$Berkman Klein Center, Harvard University, Cambridge, MA, USA \\
  \small $^6$Information School, University of Washington, Seattle, WA, USA \\
  \small $^7$Center for an Informed Public, University of Washington, Seattle, WA, USA \\
  \small $^8$Department of Logic and Philosophy of Science, UC Irvine, Irvine, CA, USA \\
  \small $^9$Center for Socially Engaged Philosophy, UC Irvine, Irvine, CA, USA \\
  \vspace{2mm} \\
  \small $^\ast$To whom correspondence should be addressed; E-mail: \texttt{jbakcoleman@gmail.com}
}

\renewcommand{\shorttitle}{Industry Influence in High-Profile Social Media Research}

\hypersetup{
pdftitle={A template for the arxiv style},
pdfsubject={q-bio.NC, q-bio.QM},
pdfauthor={Joseph Bak-Coleman, Jevin West, Cailin O'Connor and Carl Bergstrom},
pdfkeywords={Social Media, Industry Influence, Agnatology, Science of Science },
}

\begin{document}
\maketitle
\begin{abstract}
To what extent is social media research independent from industry influence? Leveraging openly available data, we show that half of the research published in top journals has disclosable ties to industry in the form of prior funding, collaboration, or employment. However, the majority of these ties go undisclosed in the published research. These trends do not arise from broad scientific engagement with industry, but rather from a select group of scientists who maintain long-lasting relationships with industry. Undisclosed ties to industry are common not just among authors, but among reviewers and academic editors during manuscript evaluation. Further, industry-tied research garners more attention within the academy, among policymakers, on social media, and in the news. Finally, we find evidence that industry ties are associated with a topical focus away from impacts of platform-scale features. Together, these findings suggest industry influence in social media research is extensive, impactful, and often opaque. Going forward there is a need to strengthen disclosure norms and implement policies to ensure the visibility of independent research, and the integrity of industry supported research.
\end{abstract}


\section*{Introduction}
The proliferation and infusion of social media into daily life has raised concerns about its wide-ranging, and at times harmful, psychological and sociological effects \cite{Lorenz-Spreen2022ADemocracy, orben2024mechanisms,valkenburg2022social,Allington20,gerbaudo2023angry, abdalla2021grey}. Scientists studying those effects are largely dependent on platforms for access to data, on-platform experiments, and even basic information about platform design and functionality \cite{Wagner2023IndependencePermission,Krause2025, bakcoleman2025risksindustryinfluencetech}. This relationship between science and industry raised fears among early computational social scientists of a future in which the field was characterized by, ``a privileged set of academic researchers presiding over private data from which they produce papers that cannot be critiqued or replicated \cite{lazer2009computational}.'' Such a scenario, the authors argued, would make it difficult to accumulate knowledge and verify findings. These fears are especially salient for research into potential harms of social media, where industry may have incentives to suppress inconvenient discoveries or redirect scientific attention away from fraught topics \cite{bakcoleman2025risksindustryinfluencetech, abdalla2021grey}.

Historically, industries producing products that cause harm have successfully redirected scientific research through various techniques \cite{Oreskes2010MerchantsWarming, proctor2012golden,OConnor2019TheSpread,Chartres2016AssociationMeta,Fabbri2018TheReview}.  At the scale of individual studies, evidence across a wide range of scientific contexts has consistently demonstrated that research linked to industry exhibits bias towards industry-preferred outcomes \cite{10.1001/jamanetworkopen.2023.43425,Lexchin1167,Oreskes2010MerchantsWarming,10.1001/jama.279.19.1566,Fabbri_Holland_Bero_2018}. Notably, this biasing effect can occur without fraudulent research or nefarious industry interference, emerging instead from simple misalignment of incentives \cite{bakcoleman2025risksindustryinfluencetech}. For example, industry actors may selectively supports scientists whose research topics and methods tend to yield findings favorable to industry, biasing the larger scientific record \cite{Holman2017ExperimentationSelection, freeborn2024industrial, bakcoleman2025risksindustryinfluencetech}. 

Despite the potential for industry influence on the study of social media, its extent of such remains poorly characterized. Here, we leverage public data to evaluate these influences in high-profile social media research. We begin by characterizing the degree to which published research is tied to industry, and whether such ties are made visible by competing interests policies. We further examine the relative impact of industry connected and independent research in terms of citations and public attention.  Next, we examine authors' ties to industry, quantifying the breadth of industry ties, the degree to which firms selectively invest in specific researchers, and the independence of high profile scientists in the field. We then examine how ties to industry enter the process of evaluating manuscripts by way of editors and peer-reviewers. Finally, we reveal topical biases in research associated with industry.  

\section*{Methods}
We queried OpenAlex \cite{Priem2022OpenAlex} to identify a core corpus of research articles on social media published in \textit{Science}, \textit{Nature}, and \textit{PNAS}, as well as their common transfer journals (i.e. \textit{PNAS Nexus}, \textit{Science Advances}, \textit{Nature Communications} and \textit{Nature Human Behavior}). We focused on these journals given their broad readership and role as hubs for interdisciplinary research on the effects of social media. Narrowing our focus to specific journals also enabled us to tailor our analyses to their specific competing interest policies and to manually validate industry ties (See SI Sec. \ref{DIIdentification}).  We searched for all articles mentioning Social Media, LinkedIn, Facebook, Instagram, Youtube, WhatsApp, or Twitter in their title or abstract, leveraging the bibliographic coupling network to identify a core corpus of research articles (See SI Sec. \ref{DataCollection} ) Our corpus comprises \dataProcessingFinalcorpussize{} articles published by \authorMetricsUniqueauthors{} unique authors.

For each author, we identified any year in which they received funding from, collaborated with employees of, or were employed by one of four technology companies that own and operate one or more social media platforms: Meta, X, Google, and Microsoft (Supplementary Information Table \ref{ditotals}). We identified these ties using data from OpenAlex as well as through industry-announced RFPs and fellowships. We manually validated every funding and employment tie using sources independent from OpenAlex. Details can be found in our supplement (Sec. \ref{DIIdentification}). We refer to these industry links as ``disclosable ties'' throughout, as it enables us to distinguish them from competing interests, which may be more expansive. 

We categorized an article as possessing a disclosable tie if at least one of the authors had a detected connection to industry within the journal's specified period for competing interests: three years for \textit{Nature}, four years for \textit{PNAS},  and five years for \textit{Science} \cite{SpringerCompetingInterests, PNASPolicies, SciencePolicies}. We note that under this definition, funding from one firm would be considered a disclosable tie regardless of whether that firms' platform is explicitly mentioned in the paper. This definition is appropriate as studies about a competitor's products would warrant disclosure, and discussions of social media generally may impact all firms. To evaluate impact, we combined our data from OpenAlex with data from Altmetrics. Finally, to evaluate whether journal editors had disclosable interests, we applied the detection methods above and further examined their publicly posted CVs when available.

Although modest in size, our corpus has had a large impact. As of November 6th, 2025, papers therein have garnered a cumulative Altmetric attention score of \totalsAltmetric{}. Within the academy, the work has been  cited \totalsDimensions{} times, averaging \totalsAvgdimensions{} citations per paper. Among the public, the literature in our corpus has been mentioned in \totalsNews{} news articles, resulting in \totalsSocialmedia{} mentions on social media. This body of work has been mentioned \totalsWikipedia{} times on Wikipedia and has been referenced in \totalsPolicy{} policy documents. The authors within the corpus are influential as well, with a cumulative 7.1 million citations. As such, the research and researchers examined here likely play an out-sized role in shaping the direction of the field, and broader understandings of social media. 

\section*{Results}
\subsection*{Literature}
\begin{figure}[htbp!]
    \centering
    \includegraphics[width=0.65\linewidth]{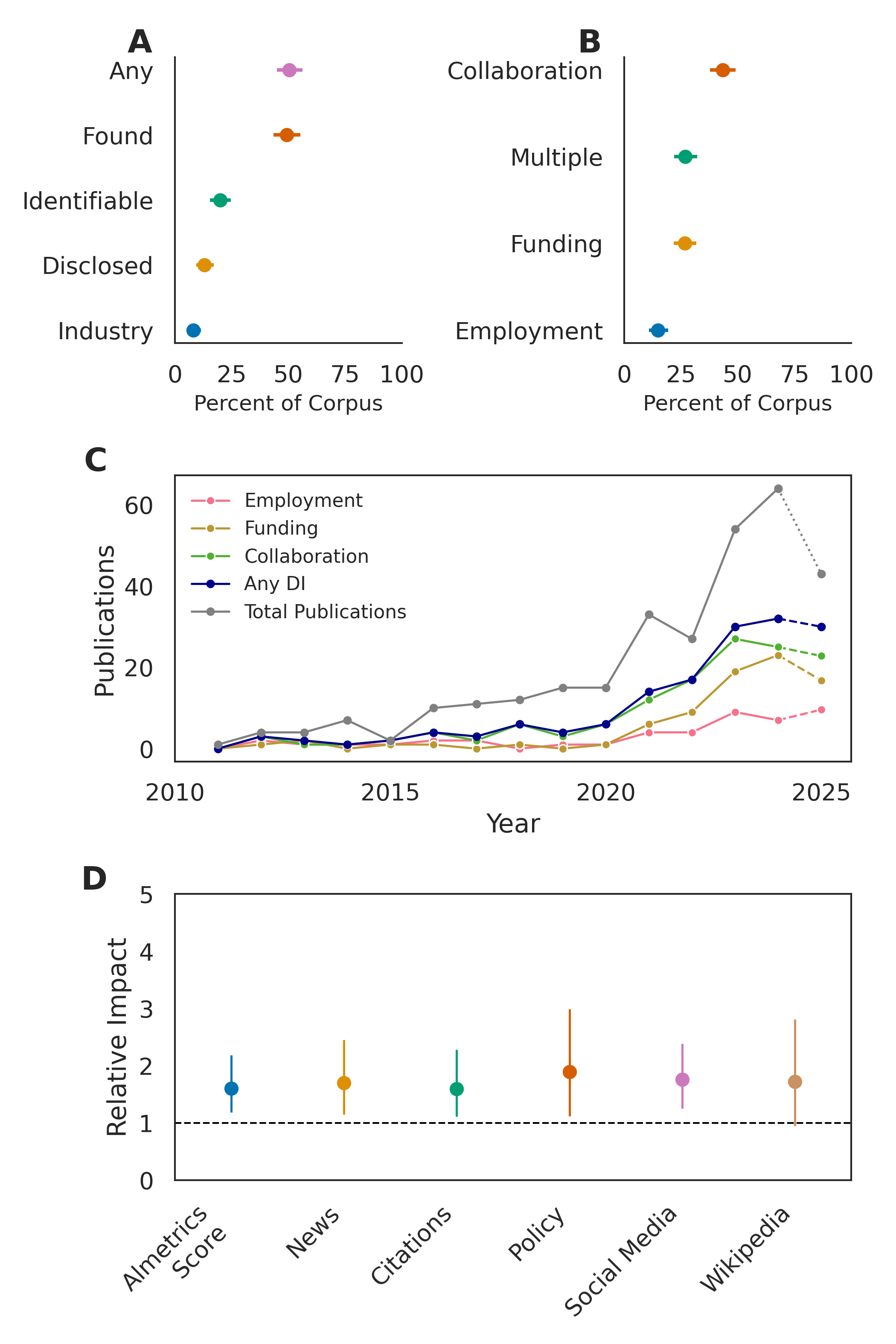}
    \caption{ Article-level descriptive statistics. \textbf{A)} Disclosable ties that were either indicated by author affiliation (Industry), indicated in competing interest statements (Disclosed), found somewhere in the manuscript (Identifiable), found through our broader search (Found), or any combination of the above (Any). \textbf{B)} Percentage of papers with various sorts of disclosable industry connections 
    \textbf{C)} Time-series of total articles with various disclosable ties
    \textbf{D)} Comparison of altmetric and citation impact. Values indicate the relative impact of papers with disclosable ties compared to research presumed to be independent. Vertical lines indicate 95\% credible regions. 
    }
    \label{fig:papers}
\end{figure}

Social media firms possess an outsize capacity for internally producing scientific research, given their access to data, sole capacity to run platform-wide experiments, and teams of scientists on staff \cite{bakcoleman2025risksindustryinfluencetech}. Consequently, research they produce alone or in collaboration with academic researchers often provides novel insights and appears in high-profile venues \cite{maccarthy2020transparency, Wagner2023IndependencePermission}. In our corpus, \literatureIndustryMean{}\% (95\% CR: [\literatureIndustryLow{}\%, \literatureIndustryHigh{}\%]) of articles indicate industry-affiliated employees among the authors (Fig. \ref{fig:papers}A). 

Competing interests statements are a key mechanism by which authors with academic employment can transparently report relevant ties to industry. Our manual review of the literature reveals that explicit declarations of competing interests are uncommon, appearing in \literatureDisclosedMean{}\% [\literatureDisclosedLow{}\%, \literatureDisclosedHigh{}\%] of the publications in our corpus (Fig. \ref{fig:papers}A). Beyond competing interest statements, some authors disclose ties to industry elsewhere such as in funding or acknowledgment statements. We refer to ties that can be identified anywhere in a published article as ``identifiable.'' Between affiliations, competing interest statements, and other locations in the manuscript \literatureIdentifiableMean{}\% [\literatureIdentifiableLow{}\%, \literatureIdentifiableHigh{}\%] of papers were identifiably linked to industry (Fig. \ref{fig:papers}A).

However, these transparency mechanisms rely entirely on voluntary disclosure by authors, raising the possibility that some ties go undisclosed. Based on public data, we found disclosable ties for  \literatureFoundMean{}\% [\literatureFoundLow{}\%, \literatureFoundHigh{}] of published work, meaning fewer than half of works with diclosable ties in fact disclosed them (Fig. \ref{fig:papers}A). Even among the \affirmnolitAnyN{} publications in our corpus which explicitly declared no competing interests, we found that \affirmnolitFoundMean{}\% [\affirmnolitFoundLow{}\%, \affirmnolitFoundHigh{}\%] nonetheless did have disclosable ties to industry (SI Fig \ref{fig:AffirmNo},  Table \ref{tab:affirmNoPaperTable}).

Competing interests vary in nature. We categorized disclosable ties into employment, collaboration, and funding. Of these types, prior collaboration was most common, followed by funding, with employment being relatively rare (Fig. \ref{fig:papers}B). Similar frequencies of connection type were observed among both research that  identified existing links to industry and papers which failed to do so. This suggests that failures to disclose are not related to norms surrounding the type of tie, such as disclosing funding yet not collaboration or employment (Fig. \ref{fig:UndisclosedTypes}). Viewed over time, our data reveal an overall increase in the volume of social media research, with an uptick in proportion of papers tied to industry beginning in 2020 (Fig, \ref{fig:papers}C). We find no clear indication that the probability of disclosure for competing interests has increased  over time (Bayesian Binomial Regression, $\beta$=\disclosuretrendBetamean{} [\disclosuretrendBetalow{}, \disclosuretrendBetahigh{}], SI Fig. \ref{norms_papers_disclosure}).

To further assess the impact of industry on both scientific and public belief, we leveraged Altmetric data, which tracks online engagement with scientific papers within and beyond the academy. This approach revealed that industry-connected research garners approximately twice the impact of independent work on average (Fig. \ref{fig:papers}D, SI Table \ref{tab:altmetricImpact}). Industry-connected work receives more citations within the academy, is more frequently referenced in policy documents, more often discussed on social media and the news, and is more likely to be referenced on Wikipedia. It is unclear the degree to which this impact is attributable to unique access to data, industry PR campaigns, characteristics of the funded researchers, or some other cause. Nevertheless, these results indicate that industry-tied research has an outsized influence on our understanding of social media.

\subsection*{Authors}
We next evaluated how these patterns of industry connection are reflected among authors in the field. A total of \authorMetricsUniqueauthors{} distinct authors were listed among the publications in our corpus. While half of publications have discernable links to industry, most individiual authors do not,  with only \authorsaAnyMean{}\% [\authorsaAnyLow{}, \authorsaAnyHigh{}] having one or more found disclosable tie (SI Table \ref{tab:author_discoverability}). In our author-scale analysis, as with the paper-level analysis, collaboration and funding were more common as sources of disclosable ties than were current or past employment (Fig \ref{fig:authors}A).  

Our analysis additionally revealed differences across firms in their breadth of engagement with the academy (Fig \ref{fig:authors}C, SI Table \ref{tab:author-firms}). Ties to Meta were most common among authors in our corpus (\authorsbMetaMean{}\% [{\authorsbMetaLow{}, \authorsbMetaHigh{}], Fig \ref{fig:authors}B). Google (\authorsbGoogleMean{}\% [{\authorsbGoogleLow{}, \authorsbGoogleHigh{}]) and Microsoft (\authorsbMicrosoftMean{}\% [{\authorsbMicrosoftLow{}, \authorsbMicrosoftHigh{}]) trailed behind, and very few authors had connections to X (\authorsbTwitterMean{} \% [\authorsbTwitterLow{}, \authorsbTwitterHigh{}]). Finally,  \authorsbMultipleMean{} \% [\authorsbMultipleLow{}, \authorsbMultipleHigh{}] of authors engaged with more than one firm.

Historically, industries seeking to influence science have invested selectively and heavily in industry-favorable researchers, or those who work on industry-favorable topics \cite{Oreskes2010MerchantsWarming, Pinto2023EpistemicBias}.  There is evidence that large technology companies, like Google, have attempted to select and groom friendly academics to push research in favorable directions \cite{abdalla2021grey}. To evaluate whether industry invests broadly or selectively, we calculated the number of years each author was funded by or collaborated with industry, summed across firms (See SI Sec. \ref{Authors}). This metric provides an estimate of the longevity and breadth of industry support an individual scientist received. Across our \authorMetricsUniqueauthors{} authors, the average company-years of support was \authorsupportyearsMean  [\authorsupportyearsMeanlow{}, \authorsupportyearsMeanhigh{}] years  with a maximum of \authorsupportyearsMax{} company-years. For this metric, the Gini coefficient is \authorsNonemployedgini{}, suggesting consistent industry collaboration and funding is highly concentrated among a small number of academics (Fig. \ref{fig:authors}C). Indeed, the top 10\% of authors account for \authorsToptenpercentcontrol{}\% of total industry investment by this metric.

\begin{figure}[htbp!]
    \centering
    \includegraphics[width=0.65\linewidth]{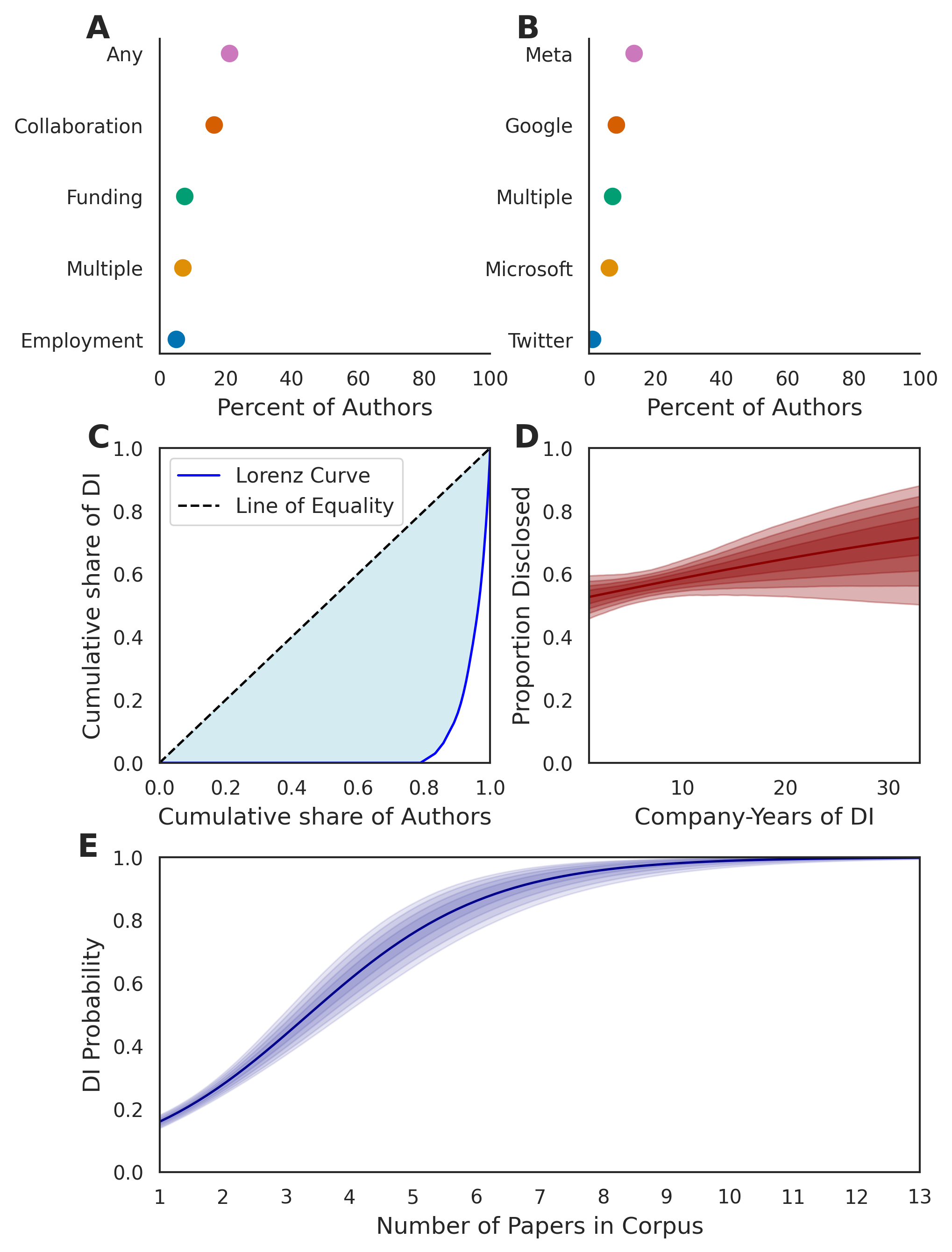}
    \caption{ Author-level descriptive statistics. \textbf{A)} The percentage of authors possessing a given type of tie to industry. \textbf{B)} The percentage of authors engaging with each firm \textbf{C)} Lorenz curve showing inequality in access to industry funding and collaboration, as measured in total years of funding and collaboration per firm \textbf{D)} Rate of disclosure as a function of the number of years in which an author was employed by, funded by, or collaborated with industry. Shaded bands correspond to the 50, 75, 89, and 95\% credible regions for the expected probability. \textbf{E)} The probability of an author having at least one disclosable tie as a function of the number of papers on social media published in high-profile journals.}
    \label{fig:authors}
\end{figure}

We found little indication that authors with longer-standing ties to industry have higher per-paper probabilities of declaring relevant ties as competing interests ($\beta=$\authordisclosureprobBetamean{} [\authordisclosureprobBetalow{}, \authordisclosureprobBetahigh{}] (Fig. \ref{fig:authors}D, SI Table \ref{tab:author_disclosure_params}). For example, ten years of industry funding and collaboration are associated with a per-paper disclosure rate of around \authordisclosureyearsTenmean{}\% [\authordisclosureyearsTenlow{}, \authordisclosureyearsTenhigh{}]: an absolute increase of \authordisclosureyearsDiffmean{}\% [\authordisclosureyearsDifflow{}, \authordisclosureyearsDiffhigh{}] above the disclosure rates of researchers with a single year's association. 

Finally, we examined the association between academic profile in the field, proxied by number of high-profile publications on social media research, and relationships to industry. A Bayesian logistic regression indicated authors with 3 or more papers in these journals are likely to have disclosable ties, and authors with six or more almost invariably do ($\beta=$\authorprolificprobBetamean{} [\authorprolificprobBetalow{}, \authorprolificprobBetahigh{}], Fig. \ref{fig:authors}E, SI Table \ref{tab:author_prolific}). As such, there appear to be few fully independent researchers whose work on social media is routinely accepted and published in top journals. 

\subsection*{Editors and Reviewers}
High-profile scientists within a discipline serve as important gatekeepers in positions as academic editors and peer reviewers. Our results above suggest high-profile authors in social media research typically have longstanding ties to one or more social media companies. To the extent that these authors' perspectives, preferences and priorities align with industry interests, their gatekeeper roles could extend industry influence to research produced by independent authors. To prevent such effects, editors and reviewers are typically required to disclose competing interests before handling a manuscript; conflicts may warrant recusal. 

\begin{figure}[htbp!]
    \centering
    \includegraphics[width=0.65\linewidth]{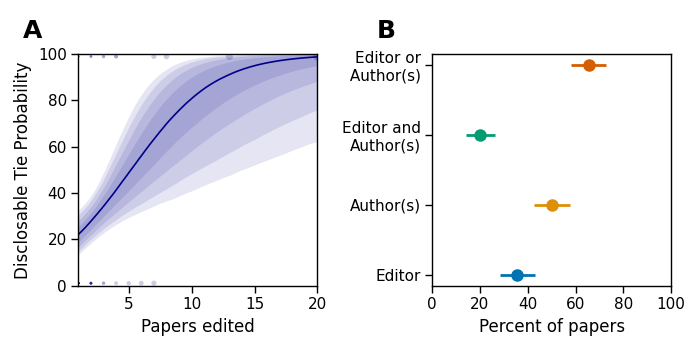}
    \caption{ \textbf{A)} The probability that an editor has a disclosable tie to industry as a function of the number of social media papers handled \textbf{B)} For papers, the probability that a paper with an academic editor has editor ties, author ties, both, or either}
    \label{fig:editors}
\end{figure}

Given the lax disclosure norms in this area, we sought to determine whether industry ties go undisclosed in evaluating manuscripts. We identified \editorbasicsagainNedtiors{} academics serving as editors for \editorbasicsEditedpapers{} manuscripts (See SI Sec. \ref{EditorsAndReviewers}). Editors were nearly twice as likely as authors in the field more broadly to have at least one disclosable tie, at \editortiesMean{}\% [\editortiesLow{}, \editortiesHigh{}]. Moreover, editors who handled a larger number of social media papers in our corpus were more likely to possess a disclosable tie to industry ($\beta$=\editortieprobabilityBetamean{} [\editortieprobabilityBetalow{},\editortieprobabilityBetahigh{}], Fig. \ref{fig:editors}A, SI Table \ref{tab:editor_params}). Overall, a third (\editedpapertiesMean{}\% [\editedpapertiesLow, \editedpapertiesHigh{}]) of papers with academic editors in our corpus were handled by an editor with ties to industry that occurred within the journals' time-frame for disclosure. None these ties were formally disclosed in the published research. 

Combined with our data on authors, we find that only a minority of research published (\editorialindependenceMean{}\% [\editorialindependenceHigh{}, \editorialindependenceLow{}]) is entirely free from disclosable ties by authors or editors, where as the majority (\editorauthtiedMean{}\% [\editorauthtiedLow{}, \editorauthtiedHigh{}]) possess  editorial or authorial disclosable ties, or both (Fig. \ref{fig:editors}B, table to add). 

Beyond ties to industry, editors are expected to recuse themselves from handling collaborators' research. In our dataset, however, 
 \editorcoauthorPmean{}\% [\editorcoauthorPlow{}, \editorcoauthorPhigh{}] of papers handled by industry-tied editors were written by their ongoing (i.e., published the same-year) or recent (i.e., within four years) co-authors. Handling of co-author manuscripts appears to be exclusive to industry-tied editors ($\beta= $\editorcoauthorBetamean{}  [\editorcoauthorBetalow{}, \editorcoauthorBetahigh{}]). We identified no examples of independent editors handling their co-authors' research. 

Our data on the literature and authors relied solely on identification of funding ties through past disclosures in published research as well as public annoucements of awards from two platforms (Meta and Google). This approach is conservative; some academics may receive funding that is neither made public by companies nor disclosed in resultant publications. Our data on editors, however, is supplemented with CVs, which can include funding not awarded through an RFP and which does not result in linked published work. As such, comparing the CVs to public data can provide an estimate of the degree to which received funding is indicated in public data and detectable through our approach. 

Overall, only \editormissingMean{}\% [\editormissingLow, \editormissingHigh] of years in which academic editors were funded by a given company were identifiable through published funding declarations or in public lists of recipients of industry awards. This remarkably low sensitivity suggests that the saturation we reconstruct from public data is highly conservative, and potentially based on only a fraction of total industry influence. However, the high concentration of industry support among a select set of authors may render many of these undetected ties as redundant for the purposes of our analyses above. Most of the years of funding not found in public data most were attributable to  Meta and Google (Meta: \editormissingMetamean{}\% [\editormissingMetalow{}, \editormissingMetahigh{}] and Google: \editormissingGooglemean{}\% [\editormissingGooglelow{}, \editormissingGooglehigh{}]). Only a small portion of missing years were funded by Microsoft \editormissingMicrosoftmean{}\% [\editormissingMicrosoftlow{}, \editormissingMicrosofthigh{}] and none were funded by X.  

Peer reviewer identities are not typically made public. However, some publications listed reviewers who were willing to disclose their identities in the interests of transparency. Such disclosures are uncommon in our corpus, with \reviewerTotal{} named reviewers across \reviewerUniquepapers{} papers. We were able to identify disclosable ties with reviewers' CVs, when available. Overall, \reviewerMean{}\% [\reviewerLow{}, \reviewerHigh{}] of these reviewers had sufficiently recent ties to industry to warrant disclosure per journal policies. It is unclear whether these ties were disclosed to editors during peer review, however none were disclosed in the published literature. 

Combining our findings throughout this paper we can estimate the {\em industrial saturation} of the literature---the proportion of papers with ties to industry either through authorship or evaluation process. Of the \saturationTotalpapers{} papers in our corpus, 194 have known ties via authors or editors. For the remaining, we can leverage the rates of disclosable ties among named reviewers to estimate the probability that at least one (of two, conservatively) anonymous reviewers has a disclosable tie. Doing so leads to a corpus-wide estimate of \saturationEsitimatedmean{}\%[\saturationEstimatedlow{}, \saturationEstimatedhigh{}] saturation. In other words, no more than one out of every five high profile social media publication is expected to be independent of industry influence in production and evaluation. 

\subsection*{Topical Bias}
A key risk of industry influence is the redirection of research priorities towards industry-preferred topics, methods, causal mechanisms, or perspectives \cite{freeborn2024industrial, Holman2017ExperimentationSelection}.  In the case of social media research, one worry is that industry directs attention to other potential causes of harms created by their platforms, especially by focusing on user choices rather than algorithmic or platform design effects \cite{bakcoleman2025risksindustryinfluencetech}.  To evaluate the potential for redirection of this general sort, we applied community detection to the bibliographic coupling network, and identified five communities ranging from \biasbasicsLowcommunity{} to \biasbasicsHighcommunity{} papers (SI Sec. \ref{TopicalBias}). Topical analysis indicated that each community corresponded to distinct themes in social media research which we have labeled misinformation sharing, platform dynamics, mental health, social network analysis, and political behavior (SI Table \ref{tab:communities}). 

Our results above suggest that \literatureFoundMean{}\% of published research in this area is tied to industry. In the absence of topical bias by industry, we'd anticipate similar proportions across each of these domains. For mental health, social network analysis, and political behavior, the relative abundance of independent and industry-tied work is consistent with this expectation (Fig \ref{fig:bias}A, SI Table \ref{tab:communityProbs}). We find an over-abundance of industry connection in the cluster of research focused on misinformation sharing. By contrast, industry connections are sparse in the the cluster related to platform dynamics. In a supplemental analysis we examined relative word frequencies within abstracts and found that the community with stronger ties to industry tended to focus on more on experiments related to people sharing false news headlines (SI Fig. \ref{fig:suppWordBias}). By contrast, terms emphasizing network structure and dynamics were linked to the comparatively independent topic. 

\begin{figure}[htbp!]
    \centering
    \includegraphics[width=0.8\linewidth]{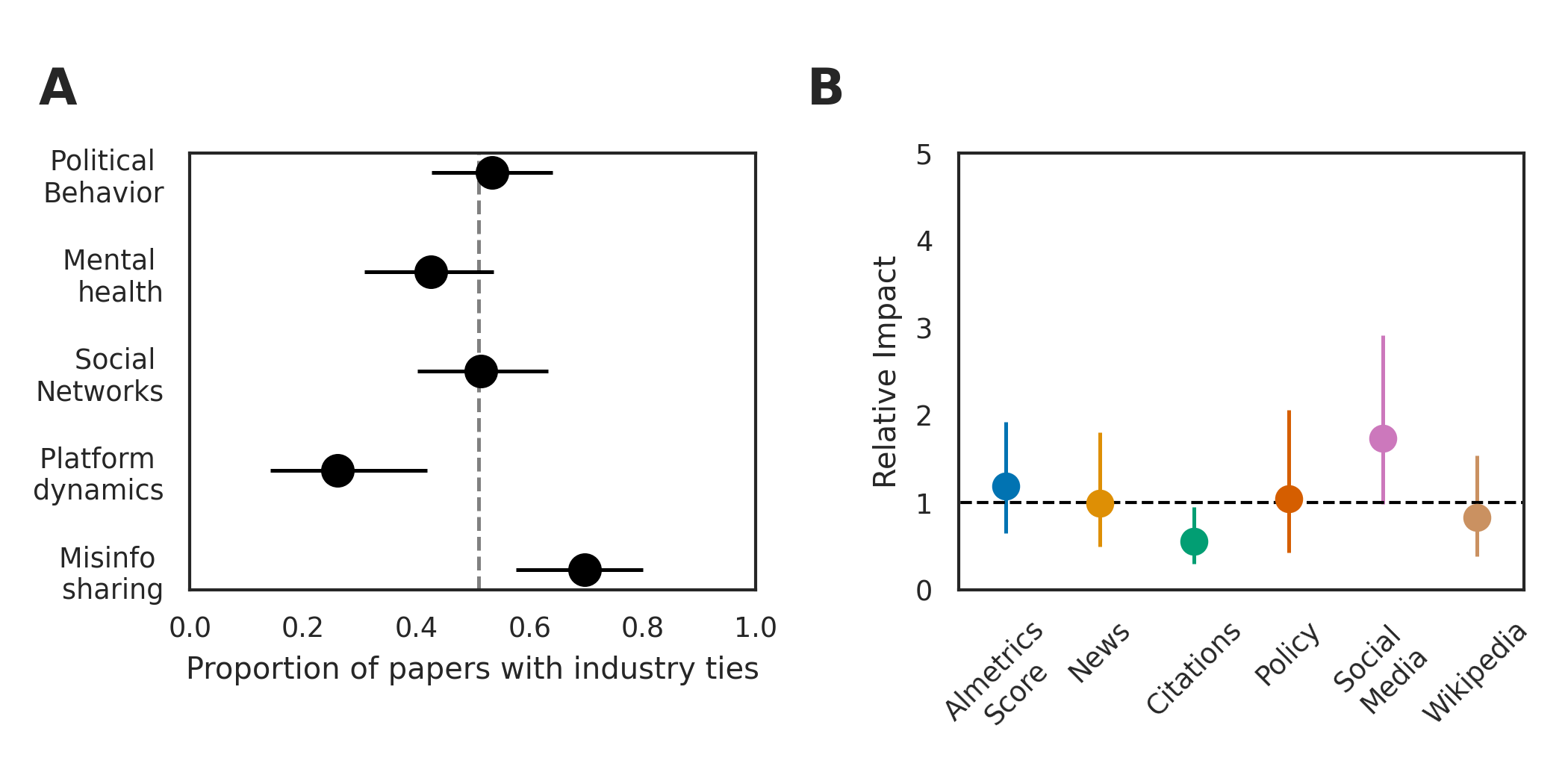}
    \caption{ \textbf{A)} Proportion of papers with ties to industry across the five topical communities identified by bibliographic coupling \textbf{B)} Dots indicate per-paper relative impact of the industry-connected misinformation sharing group and the more independent platform dynamics group. Vertical lines indicate 95\% Credible regions.}
    \label{fig:bias}
\end{figure}


While we cannot definitively show that industry funding in this area is redirecting attention from products to consumers, our results are consistent with this possibility.  It is especially notable that research into platform dynamics---which should be likely to reveal potential harms from platforms themselves---is underfunded by industry. 

Additionally, we find that work within the misinformation sharing cluster garners half as many per-paper citations as that in the platform dynamics cluster.  This suggests that this work is of lesser interest within the scientific community (Fig. \ref{fig:bias}B). However, there are nearly twice as many papers in the misinformation sharing category, possibly as a result of industry funding disparities (Table \ref{tab:altmetricImpactCommunity}).  As a result the misinformation cluster ends up with similar total number of academic citations as the platform dynamics cluster.  In addition, in part due to the volume of papers, the total sum of public engagement is considerably larger for the misinformation sharing topic (SI Table \ref{tab:communities}, \ref{tab:altmetricImpactCommunitySum}).  In other words, industry support may be translating to impact both within and without the academy for research on misinformation sharing.

\section*{Discussion}
Within high profile social media literature only one in five papers indicate any ties to industry. However, we find that half of this body of literature has discernible ties to industry. Once we consider ties from reviewers and editors as well as authors, we find precisely the opposite of what disclosure statements suggest: only one in five papers are likely to have remained independent of industry throughout the entire production and evaluation process. These findings indicate that industry influence in high-profile media research is widespread but under-disclosed. Furthermore, research linked to industry exhibits topical bias, and has outsized impact within and beyond the academy. As the field's founders feared, these patterns of influence arise from long-lasting relationships between a small group of selected scientists and social media firms. These industry-tied academics disproportionately publish in high profile venues and are represented in key gatekeeping roles as editors and peer-reviewers. 

In this social media research, weak disclosure norms and failures to adhere to existing disclosure policies not only obscure industry influence but also make it difficult to identify independent research. In other disciplines where collaboration between industry and academia is commonplace, strong disclosure norms are recognized as essential for mitigating bias, improving trust in science, and ensuring the integrity of manuscript evaluation \cite{relman1984dealing,cigarroa2018institutional,deangelis2000conflict,resnik2016ensuring, Pinto2020CommercialScience}. Disclosures also allow funding agencies to offset industry bias by targeting independent researchers, and empower editors to solicit independent evaluations of industry-tied research \cite{Pinto2023EpistemicBias}. At a minimum, our results highlight a need to correct the existing scientific literature and strengthen disclosure norms going forward \cite{abdalla2021grey}.  Authors and journals may wish to issue corrections to address failed disclosures in the existing literature. Independent authors may choose to go beyond boilerplate statements such as ``the authors declare no competing interests" and explicit describe their lack of entanglements, for example ``the authors have no competing interests in the form of funding, collaboration, or employment with any social media company."

By focusing on a modestly-sized corpus of high profile research we were able to ensure the accuracy of our data by manually verifying all industry ties. However, this approach comes with several limitations. In particular, it is unclear whether these results extend to lower profile, discipline-specific venues. In computer science, existing research suggests extensive industry influence, yet little is known for other fields \cite{bosten2025conflictspublishednlpresearch,Gizi_ski_2024, abdalla2021grey}. Another limitation is that our approach is unable to address causal questions about the mechanisms underlying the patters we have observed. It is unclear, for example, whether a scientist's high profile leads to the formation of industry ties, or vice versa. Similarly, it is not clear whether topical bias arises because industry connections shift individual scientists' research priorities, or because industry selectively supports scientists whose preexisting priorities and commitments are well aligned with their interests \cite{Pinto2023EpistemicBias}. Moreover, our analysis of bias is limited to broad topical bias. Future work is needed to provide more fine-grained analysis of bias among framings and findings, particularly within, rather than across, topics. 

Finally, we stress that our reliance on public data limits our ability to detect ties not surfaced in that data, such as unrestricted gifts, stock ownership, employment that does not produce publications, board membership, unpublished collaborations, and consulting. As such, our estimates should be viewed as a conservative lower bound of industry influence in high profile social media research. It is likely that total influence is greater than our results indicated.

\section*{Declarations}
\subsection*{Ethics Approval}
University of Washington's HSD determined that this study does not constitute research, as defined by federal and state regulations. 
 
\subsection*{LLM Use Statement}
Beyond basic orthographic and thesaural queries, generative AI was not used in drafting any part of this manuscript's text. 

\subsection*{Data Availability Statement}
While our data are drawn entirely from publicly available sources, we are not releasing the data set because it is not possible to adequately de-identify individuals represented therein.

\subsection*{Code Availability Statement}
The core of our code is sufficient to reproduce our data, raising the same concerns. To balance these concerns with transparency, we provide extensive descriptions of all statistical models sufficient to recreate them in the supplement. 

\subsection*{Competing Interests}
J.B. has engaged in paid consulting for the United Nations on topics related to social media and technology and declares no competing interests with regard to any company that owns or operates a social media platform. C.B. and J.W. attended the Social SciFoo conference in 2021 and J.W. attended Social SciFoo conference in 2023; Google and Meta, respectively, covered the accommodations for participants. In 2022, C.B. and J.W. gave a paid talk at Amazon's finance division about data reasoning. J.W. was senior personnel on a \$50k Meta grant in 2018 supporting crisis informatics in conflict zones, but he received no funding for this work. The Center for an Informed Public at UW, of which J.W. and C.B. are affiliated, has received funding and Azure credits from Microsoft. The authors declare no other competing interests in the form of funding, collaboration and employment with any firm that owns or operates a social media platform or their related philanthropic organizations.

\printbibliography

\end{document}